\documentclass[10pt,twocolumn,letterpaper]{article}

\usepackage{cvpr}              % To produce the CAMERA-READY version
\usepackage{graphicx}
\usepackage{amsmath}
\usepackage{amssymb}
\usepackage{booktabs}
\usepackage{multirow}
\usepackage[table]{xcolor}
\usepackage{placeins}
\usepackage[pagebackref,breaklinks,colorlinks]{hyperref}

\usepackage[capitalize]{cleveref}
\crefname{section}{Sec.}{Secs.}
\Crefname{section}{Section}{Sections}
\Crefname{table}{Table}{Tables}
\crefname{table}{Tab.}{Tabs.}

\def\cvprPaperID{6294} % *** Enter the CVPR Paper ID here
\def\confName{CVPR}
\def\confYear{2024}

\begin{document}

%%%%%%%%% TITLE - PLEASE UPDATE
\title{Analyzing the Effect of $k$-Space Features in MRI Classification Models}

\author{Pascal Passigan\\
MIT\\
{\tt\small ppxscal@mit.edu}
% For a paper whose authors are all at the same institution,
% omit the following lines up until the closing ``}''.
% Additional authors and addresses can be added with ``\and'',
% just like the second author.
% To save space, use either the email address or home page, not both
\and
Vayd Ramkumar\\
MIT\\
{\tt\small vayd@mit.edu}
}
\maketitle

%%%%%%%%% ABSTRACT
\begin{abstract}
The integration of Artificial Intelligence (AI) in medical diagnostics is often hindered by model opacity, where high-accuracy systems function as "black boxes" without transparent reasoning. This limitation is critical in clinical settings, where trust and reliability are paramount. To address this, we have developed an explainable AI methodology tailored for medical imaging. By employing a Convolutional Neural Network (CNN) that analyzes MRI scans across both image and frequency domains, we introduce a novel approach that incorporates Uniform Manifold Approximation and Projection (UMAP) for the visualization of latent input embeddings. This approach not only enhances early training efficiency but also deepens our understanding of how additional features impact the model’s predictions, thereby increasing interpretability and supporting more accurate and intuitive diagnostic inferences.
\end{abstract}

\section{Introduction}

The application of Convolutional Neural Networks (CNNs) in medical imaging, particularly Magnetic Resonance Imaging (MRI), has transformed diagnostic methodologies, yielding high precision in tasks such as classification and segmentation. Despite these advances, the primary challenge remains the interpretability of these models. High-performing AI systems in healthcare must address opacity to gain clinical trust by providing transparent and understandable decision-making processes.

Our research introduces an innovative approach to MRI analysis by integrating $k$-space (frequency domain) features with spatial domain representations. This dual-domain methodology not only leverages the intrinsic data richness of MRIs but also fosters model transparency through enhanced feature visualization and interpretability. We hypothesize that the incorporation of $k$-domain features will significantly refine the diagnostic capabilities of CNNs, offering a more granular understanding of neural activations within the network.

\subsection{Background}

\subsubsection{MRI and $k$-space}
Magnetic Resonance Imaging (MRI) operates fundamentally through data acquisition in the frequency domain, or $k$-space. The spatial information is encoded into frequency domain data via magnetic field gradients. This results in a frequency-encoded signal that represents various spatial locations across the tissue. 

To render these $k$-space data into clinically interpretable images, an inverse Fourier transform (IFFT) is applied, translating the complex frequency domain information back into the spatial domain. In our research, we aim to leverage these $k$-space features more directly. By integrating and manipulating $k$-space features prior to this transformation, we enhance the model’s interpretability and diagnostic granularity.

The strategic use of $k$-space in MRI analysis is not just about maintaining the fidelity of the original data, but also about enhancing the predictive power of CNN classifiers \cite{Lustig2007, Otazo2015}. This approach allows us to maintain a high level of detail and information integrity that is often lost in conventional processing methods.

\subsubsection{UMAP}

Uniform Manifold Approximation and Projection (UMAP), is a nonlinear dimensionality reduction technique, similar to t-SNE \cite{vanderMaaten2008tsne}, which is used to compress the information of high dimensional vectors into low dimensional space. UMAP is useful in that it translates the structures of high dimensional data into a format that can be intuitively understood; especially in clinical practice, where the user or patient may not be accustomed to quantitative metrics used to evaluate models. \cite{McInnes2018UMAP}. UMAP is widely used in fields like bio-informatics, particularly in single-cell genomics, where it helps in visualizing and interpreting large and complex datasets. Its ability to maintain both local and global data structures makes it superior for tasks where an understanding of relationships within the data is crucial. Moreover, UMAP's flexibility allows it to be used not only for visualization but also as a general-purpose dimensionality reduction technique for machine learning preprocessing. 

\section{Related Work}

The integration of $k$-space features in medical imaging, especially in magnetic resonance imaging (MRI), has seen considerable development thanks to the pioneering efforts of researchers like Lustig et al. (2007) and Otazo et al. (2015), who have demonstrated the substantial benefits of manipulating $k$-space to improve image quality and accelerate data acquisition using compressed sensing techniques \cite{Lustig2007, Otazo2015}.

Building upon these foundational advances, subsequent research has explored deep learning approaches to enhance MRI reconstructions. Zhang et al. (2019) and Han et al. (2020) integrated $k$-space data into convolutional neural networks, significantly improving image quality and diagnostic accuracy \cite{Zhang2019, Han2020}. Choi et al. (2022) utilized generative adversarial networks (GANs) to synthesize MRI images from $k$-space data, further expanding the potential applications of this data in medical imaging \cite{Choi2022}.

Moreover, innovative applications in data visualization such as Uniform Manifold Approximation and Projection (UMAP) have transformed how high-dimensional data can be interpreted. UMAP, developed by McInnes et al. (2018), has proven versatile across various domains, including biomedical imaging and genetics, where it has been used to clarify complex datasets \cite{McInnes2018UMAP}. Becht et al. (2019) and Thompson et al. (2021) have utilized UMAP to reveal intricate data patterns, demonstrating its utility in diverse scientific fields \cite{Becht2019, Thompson2021}.

Recent studies by Kim et al. (2023) and Lee et al. (2024) further illustrate the evolving role of $k$-space in medical diagnostics. Kim et al. (2023) developed a novel algorithm that enhances the reconstruction of $k$-space data for more accurate brain imaging, while Lee et al. (2024) explored the application of machine learning techniques to predict and analyze patient outcomes based on $k$-space patterns, signifying a trend towards predictive diagnostics in radiology \cite{Kim2023, Lee2024}.

Our research extends these innovations by combining advanced $k$-space data handling with UMAP's visualization capabilities. This approach not only advances diagnostic accuracy but also makes MRI data more interpretable and actionable for clinical settings, where understanding and utilizing the technology can significantly impact patient care.

Moreover, building on the frequency domain data augmentation concepts by Park et al. (2021), we employ sophisticated mathematical manipulations that preserve the detailed information of the original $k$-space data, offering a novel perspective in MRI data processing that could lead to earlier and more precise diagnoses \cite{Park2021}.

\section{Methods}
\subsection{Dataset Curation}
Our dataset consists of preprocessed MRI images classified into four diagnostic categories: Mild Demented, Moderate Demented, Non Demented, and Very Mild Demented. These were sourced from \cite{sachin_kumar_sourabh_shastri_2022, kaggle_alzheimer_mri}. It has $N = 8131$ samples. 

\begin{figure}[htbp]
\centering % Centers the figure
\includegraphics[width=0.99\columnwidth]{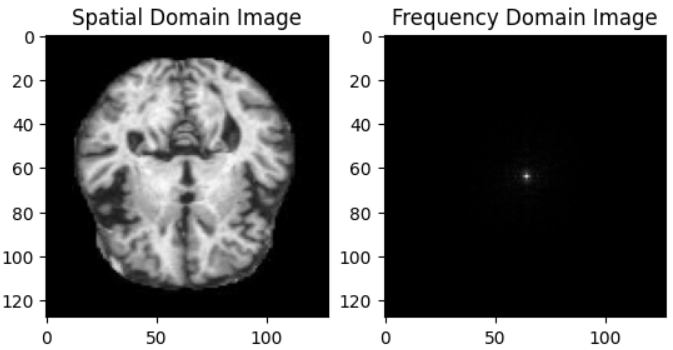} % Adjusts the size of the image to 90% of the column width
\caption{(left) One sample from the dataset, and (right) its Fourier transform.} % Caption for the figure
\label{fig:image} % Label for referencing the figure in1 text
\end{figure}

\subsection{Fourier Transform}
We apply the Fourier transform (via the FFT algorithm) to convert each image into $k$-space data, enabling the incorporation of frequency domain features, and concatenate those to the original data tensor. 

\subsection{Model Architecture}

\begin{figure}[htbp]
\centering % Centers the figure
\includegraphics[width=.9\columnwidth]{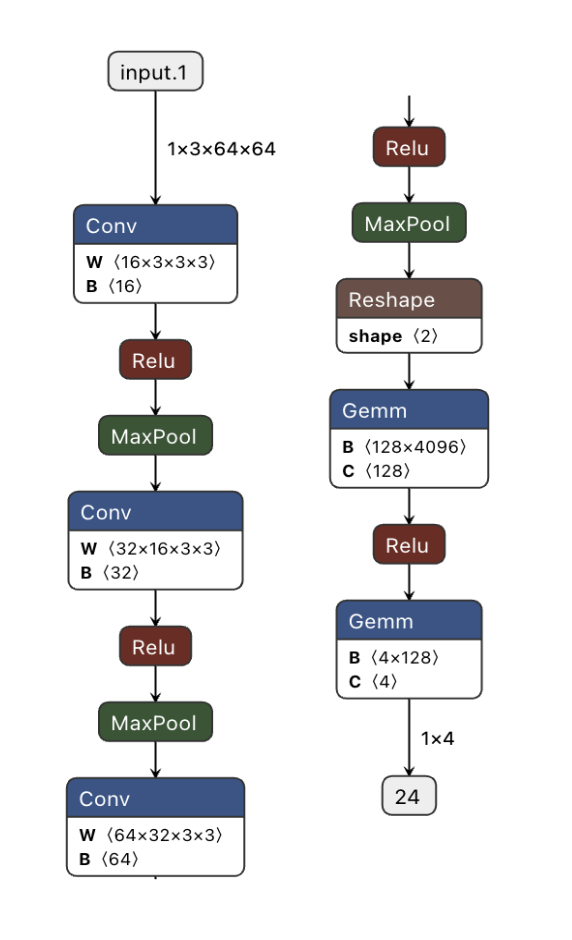}
\caption{Backbone Model Architecture} % Caption for the figure
\label{fig:model_arch} % Label for referencing the figure in text
\end{figure}

The strategy we employed in developing our model architecture lies in its first convolutional layer, which processes three distinct channels: one for spatial domain images and two for the real and imaginary components of the Fourier-transformed data, contrasting sharply with the control model that processes only grayscale image data. This allows our model to exploit the full spectrum of information contained within MRI frequency domain data, enhancing both accuracy and interoperability. A diagram of the model architecture can be found in Figure \ref{fig:model_arch}.

\subsection{Model Evaluation}

We evaluate the efficacy of our frequency-enhanced CNN classifier by comparing it to a baseline image-domain CNN, which achieves an accuracy of 99\%. Our assessment extends beyond just accuracy, considering the structure and distribution of data points in UMAP visualizations. This approach allows us to examine the spatial and cluster-based differences between the models, providing insights into how frequency domain features influence the overall model performance.

We further scrutinize the models through their confusion matrices, which offer a detailed view of classification performance across different sample types. This comparative analysis not only quantifies the improvements brought by integrating frequency domain data but also highlights specific areas where our model excels or requires further optimization.

\section{Results}

We trained both the frequency-augmented (Experimental) and the baseline CNN (Control) models five times, averaging the statistics to ensure robustness in our findings. The results, summarized in Table \ref{tab:model_performance}, suggest that the inclusion of $k$-domain data offers no significant disadvantage to the already robust base model.

\begin{table}[htbp]
\centering
\caption{Summary of Model Performances at Key Epochs}
\label{tab:model_performance}
\begin{tabular}{c c c c c}
\hline
\textbf{Model} & \textbf{Epoch} & \textbf{Val. Acc.} & \textbf{Spec.} & \textbf{AUC} \\
\hline
\multirow{3}{*}{Exp.} & 3 & 0.902 & 0.960 & 0.990 \\
                      & 6 & 0.997 & 1.000 & 1.000 \\
                      & 9 & 0.998 & 1.000 & 1.000 \\
\hline
\multirow{3}{*}{Ctrl.} & 3 & 0.922 & 0.960 & 1.000 \\
                       & 6 & 0.992 & 0.990 & 1.000 \\
                       & 9 & 0.998 & 1.000 & 1.000 \\
\hline
\end{tabular}
\end{table}

\begin{figure}[!htbp]
\centering % Centers the figure
\includegraphics[width=1 \columnwidth]{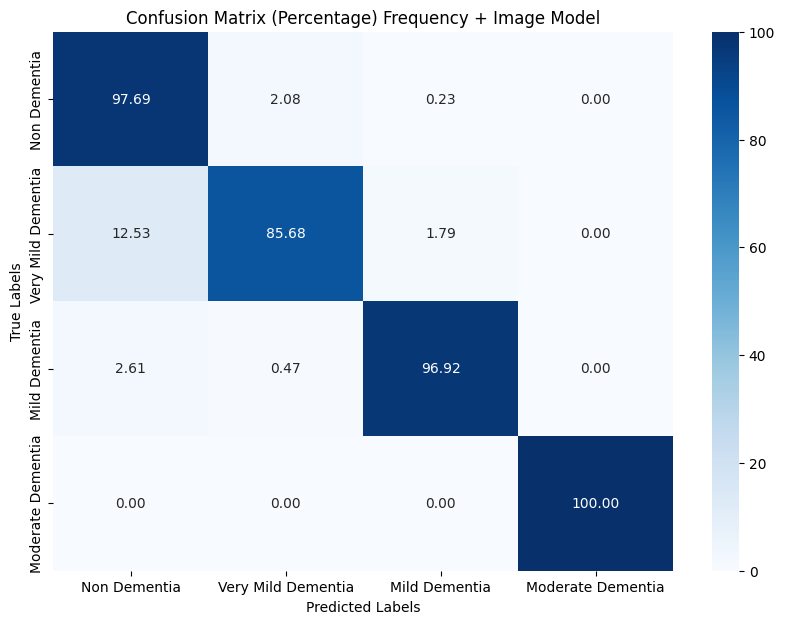}
\caption{Frequency-Augemented Model Confusion Matrix}
\label{fig:exp_conf} % Label for referencing the figure in1 text
\end{figure}

\begin{figure}[!htbp]
\centering % Centers the figure
\includegraphics[width=1 \columnwidth]{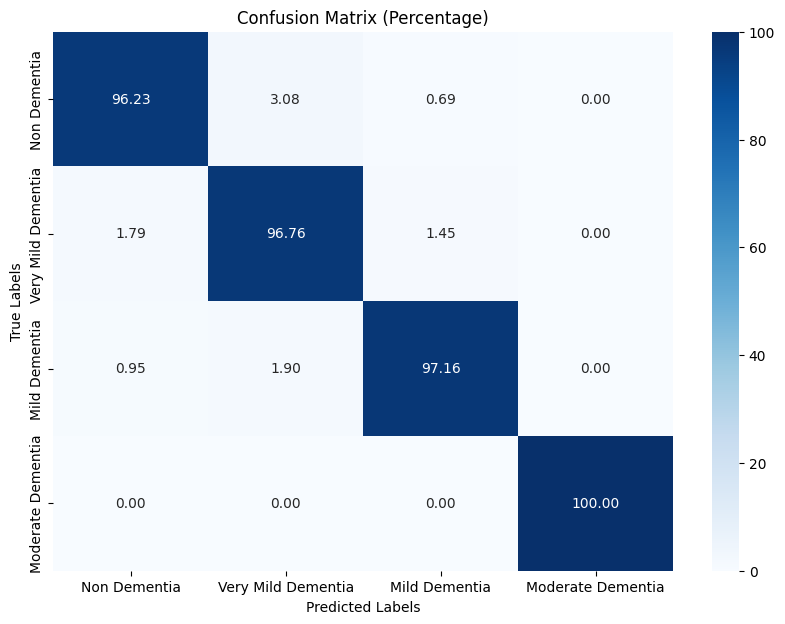}
\caption{Backbone Model Confusion Matrix} % Caption for the figure
\label{fig:base_conf} % Label for referencing the figure in1 text
\end{figure}

\begin{figure*}
  \centering

  \begin{subfigure}{0.495 \linewidth}
    \includegraphics[width=1 \columnwidth]{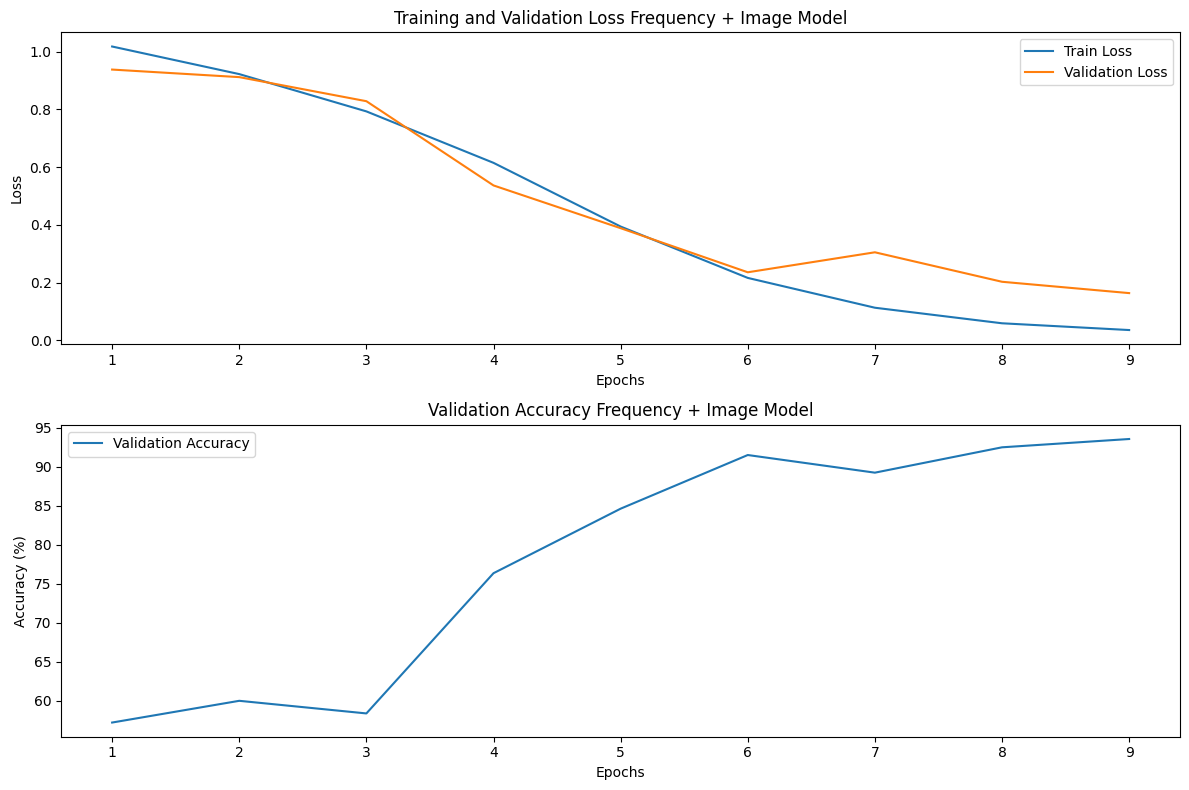}
    \label{fig:short-a}
  \end{subfigure}
  \hfill
  \begin{subfigure}{0.495 \linewidth}
    \includegraphics[width=1 \columnwidth]{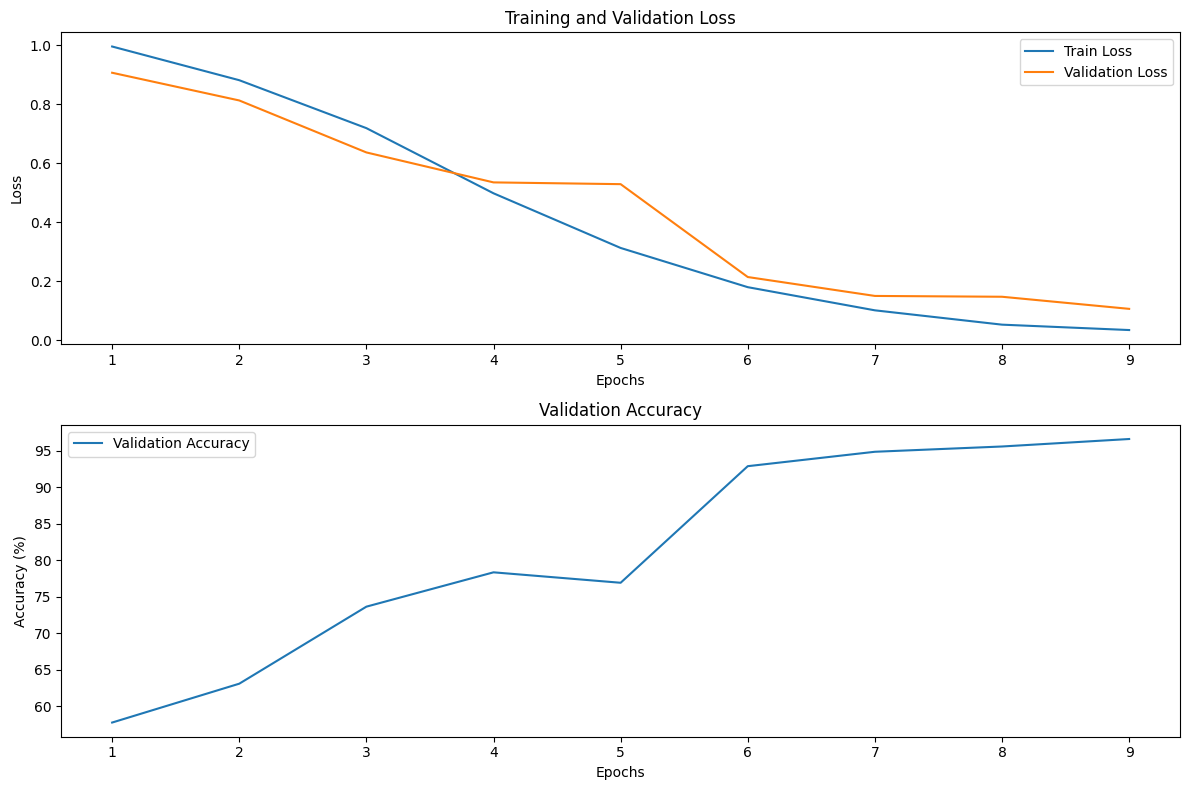}
    \label{fig:short-b}
  \end{subfigure}
  \caption{Training and Validation Loss and Accuracies for the Frequency-Augmented and Control models}
  \label{fig:lossplots}
\end{figure*}

\begin{figure*}[!ht]

\centering
\includegraphics[width=\textwidth]{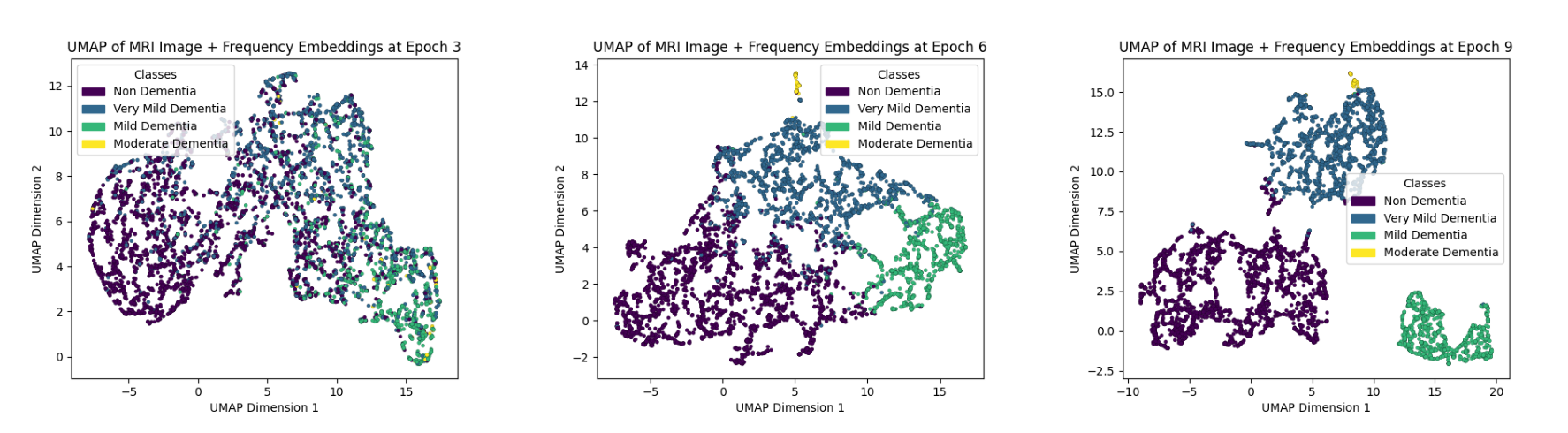}
\caption{UMAP Output of Frequency Augmented Model}
\label{fig:umap_out}
\end{figure*}

\begin{figure*}[!ht]

\centering
\includegraphics[width=\textwidth]{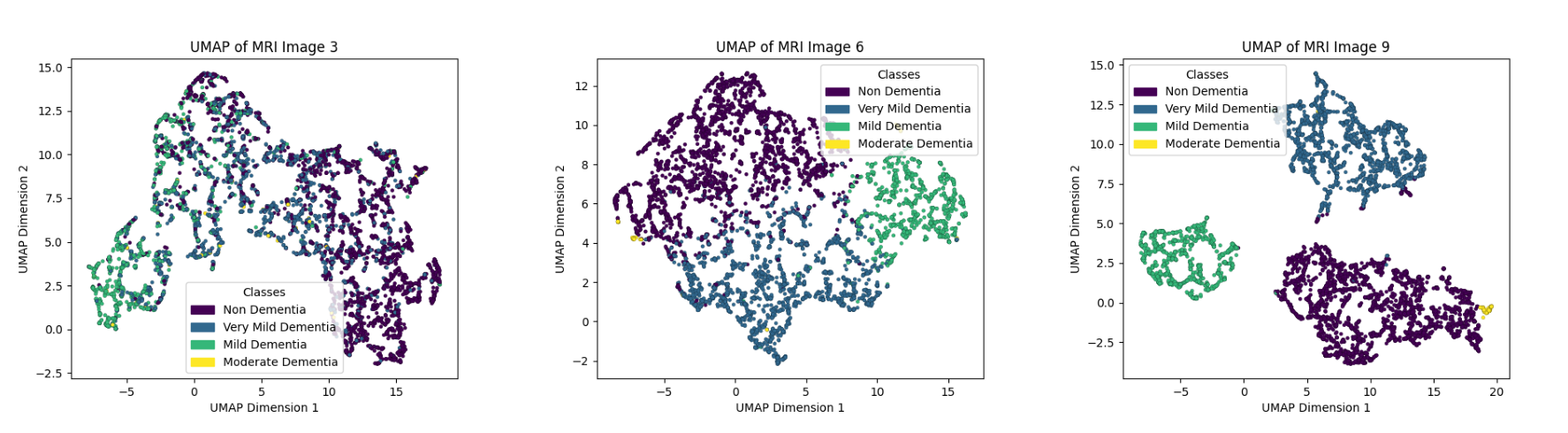}
\caption{UMAP Output of Control Model}
\label{fig:control_out}
\end{figure*}

\FloatBarrier % Ensure all figures are placed before the Discussion section

Further analysis using confusion matrices revealed differences in false positive rates between the models. While the experimental model showed improvements in reducing false positives, it struggled with distinguishing Non Dementia from Dementia more than the control model. Figures \ref{fig:exp_conf} and \ref{fig:base_conf} depict the confusion matrices for both models.

We also compared training dynamics and UMAP outputs, illustrated in Figures \ref{fig:lossplots}, \ref{fig:umap_out}, and \ref{fig:control_out}. The experimental model learned faster, particularly between epochs 3 to 6, and demonstrated a more defined cluster separation in UMAP visualizations, suggesting a hierarchical learning strategy that more effectively grouped related diagnostic categories.

\section{Discussion}

The introduction of $k$-domain features to MRI CNN classification models has shown mixed results. Early training benefits include reduced false positives and enhanced model responsiveness. However, as training progresses, these features contribute to confusion in classifying between Non Dementia and Very Mild Dementia cases. This effect is confirmed through UMAP visualizations, which depict considerable overlap between these two diagnostic categories. Such visualizations are crucial as they offer a direct view into the model's decision-making process, providing valuable insights for clinical diagnostics.

Despite these challenges, the frequency-augmented model displayed more distinct cluster formations than the control model in UMAP outputs. Notably, clusters in the augmented model exhibited dynamic movements, suggesting underlying data interactions that mimic gravitational effects around a central point. This unexpected behavior warrants further investigation as it may reveal new insights into how neural networks process and represent complex information.

UMAP's utility extends beyond traditional analysis, offering radiologists a novel application by visually mapping MRI scans within a diagnostic spectrum. This method enhances the interpretability of diagnostic outcomes, presenting data in a format that is intuitively understandable and clinically relevant.

\section{Conclusion}

Our study assessed the impact of integrating $k$-domain features into CNNs for MRI analysis. The results indicate that while these features improve certain aspects of model performance, they also introduce significant classification challenges, particularly in advanced training stages. The addition of $k$-domain features complicates the model's ability to distinguish closely related diagnostic categories, as evidenced by the increased overlap in UMAP visualizations.

Future work should focus on analyzing the dynamic clustering phenomena observed in the frequency-augmented models and refining the integration of $k$-domain features to better balance model accuracy and interpretability. Continuing this line of research could enhance our understanding of the optimal use of complex data features in medical imaging and lead to more reliable diagnostic tools in clinical settings.

While the integration of $k$-domain features presents promising avenues for enhancing MRI classification models, careful implementation is required to fully realize their benefits without compromising the clarity and accuracy of the model's outputs. The insights gained from UMAP visualizations are particularly promising, suggesting that even more nuanced interpretations of model behavior are possible and clinically applicable.

% \section{Individual Contributions}

% Pascal: Executed training runs with UMAP output and confusion matrices assessment. 

% Vayd: Did Fourier transform preprossessing, Specificity, and AUC computations.

%%%%%%%%% REFERENCES
{\small
\bibliographystyle{ieee_fullname}
\bibliography{egbib}
}
\end{document}